# Topological Insulators-Based Magnetic Heterostructure

Qi Yao[1,2], Yuchen Ji[2], Peng Chen[3], Qing-Lin He[4*], and Xufeng Kou[1,3*]

[1]ShanghaiTech Laboratory for Topological Physics, ShanghaiTech University, Shanghai 200031, China

[2]School of Physical Science and Technology, ShanghaiTech University, Shanghai, 201210, China

[3]School of Information Science and Technology, ShanghaiTech University, Shanghai, 201210, China

[4]International Center for Quantum Materials, School of Physics, Peking University, Beijing 100871, China

* Corresponding author. Email: qlhe@pku.edu.cn; kouxf@shanghaitech.edu.cn

**Abstract**

The combination of magnetism and topology in magnetic topological insulators (MTIs) has led to unprecedented advancements of time reversal symmetry-breaking topological quantum physics in the past decade. Compared with the uniform films, the MTI heterostructures provide a better framework to manipulate the spin-orbit coupling and spin properties. In this review, we summarize the fundamental mechanisms related to the physical orders host in $(Bi,Sb)_2(Te,Se)_3$-based hybrid systems. Besides, we provide an assessment on the general strategies to enhance the magnetic coupling and spin-orbit torque strength through different structural engineering approaches and effective interfacial interactions. Finally, we offer an outlook of MTI heterostructures-based spintronics applications, particularly in view of their feasibility to achieve room-temperature operation.

**Keywords:** topological insulators, magnetic heterostructures, spin-orbit coupling, structural engineering, quantum anomalous Hall effect, spin-orbit torque

## Introduction

As a new condensed matter, topological insulators (TIs) host robust topologically protected surface states accompanied with insulating bulk states. In parallel with the pursuit of the massless Dirac fermions, it is of equal significance to break the time-reversal symmetry (TRS) of the topological surfaces by introducing the perpendicular ferromagnetic order. As a result, the presence of the strong spin-orbit coupling (SOC) and exchange gap in the magnetic topological system not only leads to a variety of TRS-breaking physics [1-8], but also offers opportunities for a new generation of spintronics devices in which the highly-efficient electric control of spin states arises from the interaction between the topological and the magnetic properties of the material [9, 10].

To enable magnetic topological insulators (MTIs), one effective method is to incorporate magnetic atoms into the TI matrix. The *s* or *p* orbitals of the TI band electrons, which contribute to the topological surface states, can strongly align with the 3*d* or 4*f* orbitals of specific transition metal atoms (e.g., Mn, Cr, V) through appropriate exchange coupling interaction [11-14]. Accordingly, the magnetically-doped TI thin films have been successfully grown by molecular beam epitaxy (MBE), and they have led to the demonstrations of scale-invariant quantum anomalous Hall (QAH) effect, axion insulator state, and topological magnetoelectric effect in recent years [8, 15-17]. However, the relatively low Curie temperature ($T_C$) and local inhomogeneity-induced surface gap variation are found to be the critical bottlenecks that limit the realization of the aforementioned phenomena at deep cryogenic temperatures [7].

Alternatively, we can combine TIs with high-temperature ferro-/antiferro-magnetic materials to construct TI-based magnetic heterostructures and super-lattices. In such hybrid system, high magnetic ordering temperature can be introduced through versatile interfacial magnetic proximity effects and inter/intra-layer exchange couplings. Moreover, the structural engineering would provide an additional degree of freedom to allow the independent optimization of both the electronic and magnetic properties.

In this paper, we will review current progress in the field of MTI heterostructures in terms of TRS-breaking topological quantum physics and spin-orbit applications (Fig. 1). In Section 1, we will outline the roadmap to obtain robust QAH and axion states in modulation-doped MTI systems; in Section 2, we will discuss about the material integration strategies to achieve high-$T_C$ magnetic heterostructures; in Section 3, we will further present the combination of the magnetic order and SOC strength to explore spin-orbit torque (SOT)-related phenomena. Finally, we will briefly summarize the prospects to realize multi-functional MTI-based spintronic applications.

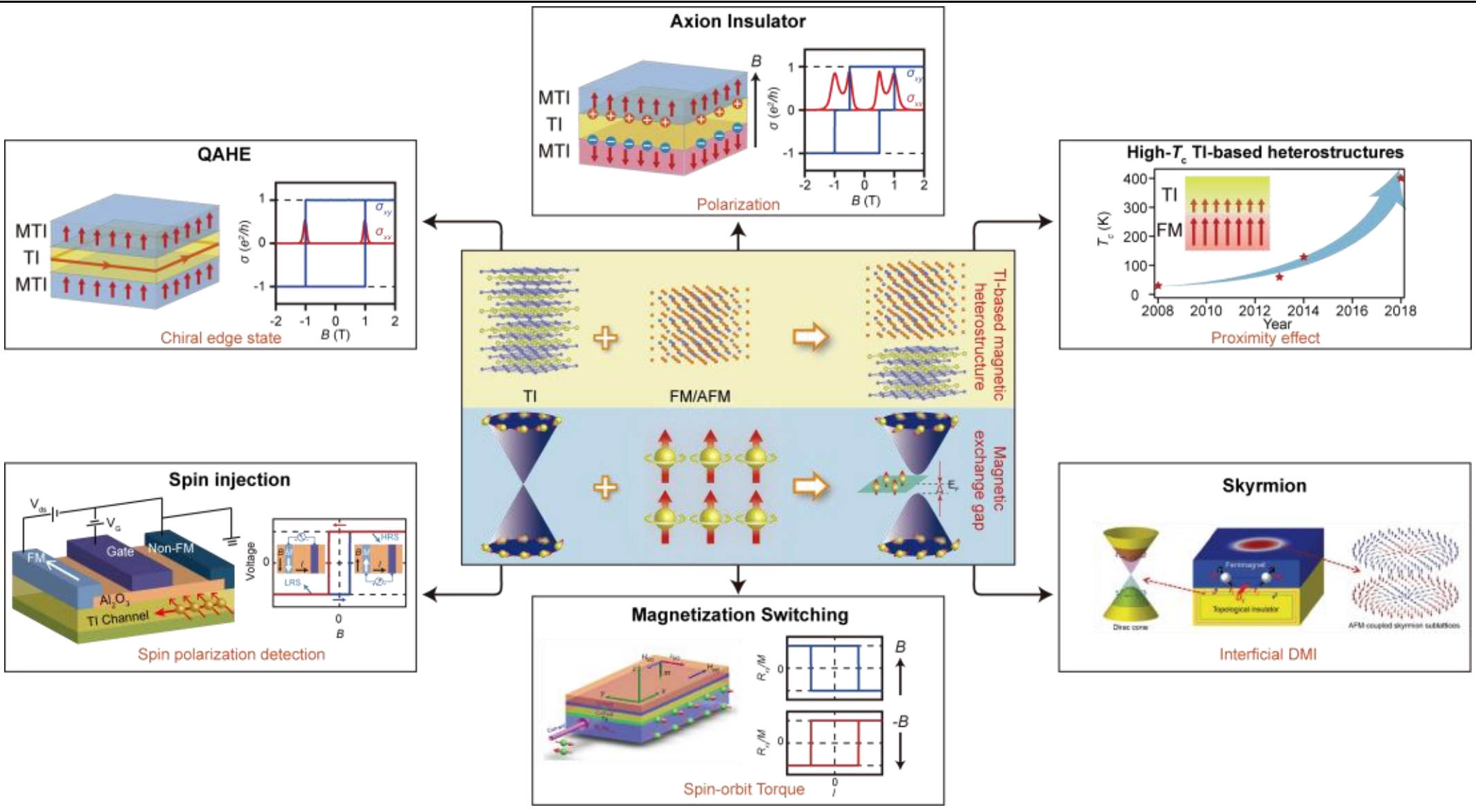

**Figure 1 TI-based magnetic heterostructures and TRS-breaking physics.**

## 1. Quantum Anomalous Hall Effect and Topological Quasi-Particles in Modulation-Doped Magnetic Topological Insulators

Ever since the discovery of the quantum Hall effect (QHE) in high-mobility two-dimensional electron gas (2DEG) systems [18], the search for its zero-field version, that is to realize dissipationless chiral edge conduction without the assistance of external magnetic field, has attracted sustaining attention for decades [15, 19-21]. In general, the role of magnetic field can be replaced by spontaneous magnetization, and the intrinsic spin-orbit interaction between charge current and magnetic moments can also give rise to the quantized Hall plateau (i.e., so-called quantum anomalous Hall effect) [22-24]. However, although the first theoretical model to realize QAH was proposed by Haldane in 1988 [23], it was not until the discovery of TIs that make the search for such suitable non-zero Chern insulators become practical.

According to four-band model of TIs, the introduction of perpendicular magnetic order would lift the spin degeneracy of the surface massive Dirac fermions [25]. When the spin splitting is well adjusted with respect to the intrinsic SOC strength, the constituent magnetic exchange field would cause band inversion in one set of the spin sub-bands while maintaining the other one in the topologically trivial regime, as shown in Fig. 2a. Experimentally, the MBE-grown Cr/V-doped $(Bi_xSb_{1-x})_2Te_3$ thin films serve as the first material platform to realize the QAH effect [4, 5]. In such uniform MTI systems, the Dirac point and Fermi level position can be tuned within the energy gap of the surface states by appropriately adjusting the $Bi_2Te_3$-to-$Sb_2Te_3$ ratio [4, 5, 26]. More importantly, the mixing between the inverted Bi/Sb $p_{1z}$-state and the Te $p_{2z}$-state enables large spin susceptibility of the band electrons to align the Cr/V *d*-orbit moments via the Van Vleck mechanism, and the long-range ferromagnetic order can be established even without the presence of itinerant carriers [1]. Consequently, when the bulk conduction is totally suppressed in the Cr/V-doped $(Bi_xSb_{1-x})_2Te_3$ at cryogenic temperatures, the Hall resistance $R_{xy}$ reaches the predicted quantized value of $\pm$ h/e$^2$, accompanied by an almost vanishing longitudinal resistance $R_{xx}$ at zero magnetic field. It is noted that in the QAH regime, the electrons can only flow along one direction with the chiral edge conduction direction determined by the magnetization orientation, as shown in Fig. 2b [4]. When the external magnetic field is applied across the coercive fields, the magnetization of the sample (i.e., edge conduction channel) is switched, hence leading to the sign change of the quantized $R_{xy}$ value (Fig. 2c).

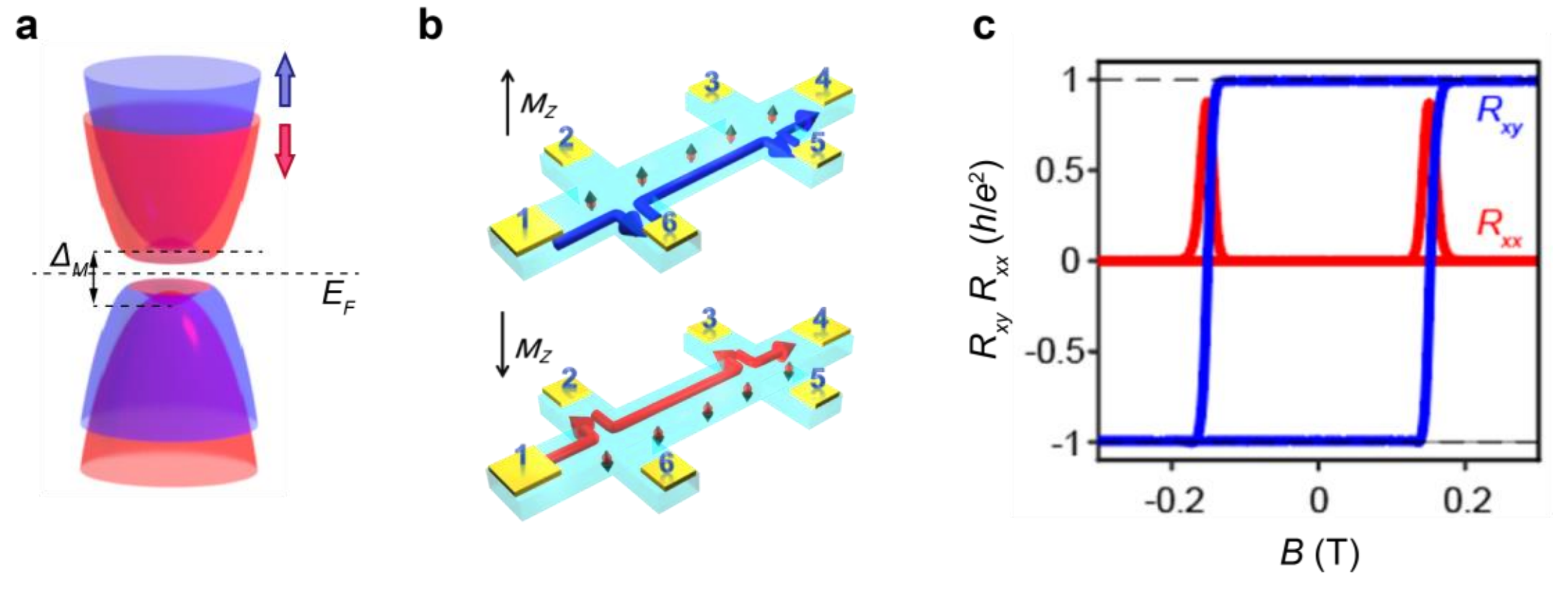

**Figure 2 Quantum anomalous Hall effect in magnetically-doped $(Bi_xSb_{1-x})_2Te_3$ thin films. a**. Schematic of the massive Dirac fermions of the surface states with spin splitting in MTIs. **b**. Illustration of the scale-invariant chiral edge conduction at the QAH state. Reproduced from [15]. **c**. Experimental observation of QAH effect in a 6 QL $(Cr_{0.2}Bi_{0.18}Sb_{0.62})_2Te_3$ sample at 50 mK.

With optimized growth condition and dedicated fabrication process, the QAH effect has been observed by several groups since 2013, yet the realization of the quantized QAH state rarely exceeded 1 K [4, 5, 15, 26-28]. Given the non-equilibrium nature of the MBE growth, it is confirmed that the inhomogeneous distribution of the incorporated magnetic dopants would result in spatial fluctuation of magnetic exchange gap [7, 29-33]. Under such circumstances, the local minimum surface gap severely limits the onset temperature of QAH (even though the mean-field gap size is calculated to be ~50 meV [29, 34]). Moreover, it is difficult to co-optimize the magnetic and topological orders simultaneously in the magnetically-doped TI systems. For instance, increasing the magnetic doping level is regarded as one possible way to enlarge the magnetic exchange gap. However, this method would not only introduce more magnetic defects (i.e., bulk dissipative conduction channels), but also dramatically reduce the overall SOC strength as more Bi/Sb atoms are substituted by the Cr/V dopants. In the extreme high-doping case, the band inversion condition may no longer be warranted, and the whole system will be turned into a trivial band insulator [35, 36].

Instead of the uniform doping strategy, advanced structural engineering makes it possible to tailor the surface states and magnetic order independently. It is known that the unique advantage of MBE growth lies in the accurate control of the doping profile along the epitaxial growth direction [37, 38]. In contrast to the Cr/V-doped $(Bi_xSb_{1-x})_2Te_3$ counterparts, the modulation-doped MTI structures, namely to selectively introduce magnetic dopants in particular regions, are demonstrated to effectively alleviate the conflicting issues mentioned above [39]. Specifically, in the proposed tri-layer structure illustrated in Fig. 3b, only the top and bottom surfaces are heavily-doped by Cr atoms [40]. Such high doping level can directly enhance the magnetic exchange coupling, hence giving rising to a larger surface gap with stabilized magnetic domains. Meanwhile, the bulk 6 QL $(Bi_{0.22}Sb_{0.78})_2Te_3$ layer remains to be intact so that the minimized dissipative bulk conduction channel with lowest defect density is well-maintained in the system. Subsequently, this tri-layer sample exhibits a more robust QAH state with higher observable temperature [40]. Likewise, the measured QAH activation gap (~80 meV) in a similar tri-layer structure is found to be more than four times larger than the uniform MTI counterpart at zero field, and its magnitude is almost insensitive to external magnetic field [41]. Furthermore, by applying the remote doping method to design the TI-MTI-TI-MTI-TI penta-layer structure as displayed in Fig. 3c, the embedded magnetic doping layer helps to enlarge the size of effective surface gap with improved spatial homogeneity while magnetic disorder-induced surface scattering is reduced [40]. As a result, relatively perfect quantization of $R_{xy}$ persists up to $T = 2$ K (Fig. 2e), which benchmarks the highest QAH temperature of the $(Bi_xSb_{1-x})_2Te_3$-based MTI systems.

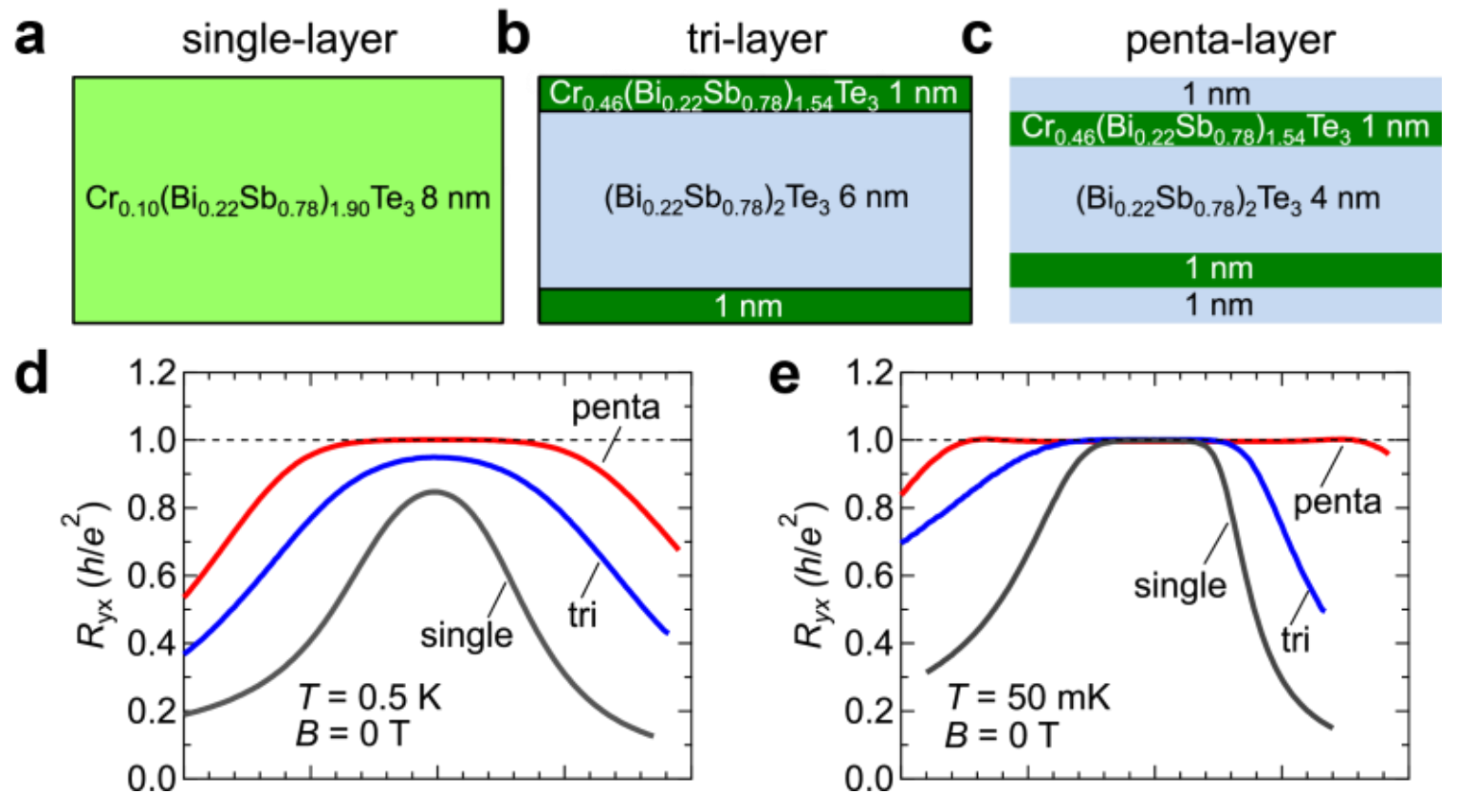

**Figure 3 Enhancement of the QAH temperature in modulation-doped MTI**

**heterostructure.** Schematic thin film structures of **a**. uniformly-doped single-layer, **b**. MTI-TI-MTI tri-layer, and **c**. TI-MTI-TI-MTI-TI penta-layer. **d**. Gate voltage dependence of Hall resistance of the single-layer (gray line), the tri-layer (blue line), and the penta-layer (red line) at 0.5 K with no external magnetic field. **e**. Nearly quantized Hall resistance (~0.97 $h/e^2$) of the penta-layer sample at 2 K. Reproduced from [40].

Following the same concept, modulation-doped growth procedure can be used to incorporate multiple dopants into one system and create exotic quasi-particles. It is found that in the QAH state, the coercivity fields of the V- and Cr-doped $(Bi_xSb_{1-x})_2Te_3$ films are $H_{C1}$ ~ 1 T and $H_{C2}$ ~ 0.15 T, respectively (Fig. 4a). Taking the advantage of such magnetic distinction, if a tri-layer V-doped $(Bi,Sb)_2Te_3$/TI/Cr-doped $(Bi,Sb)_2Te_3$ sandwich heterostructure is fabricated, an anti-parallel magnetization alignment may appear when the external magnetic field is applied between [$H_{C2}$, $H_{C1}$]. In this regard, the system is expected to transit from the QAHE state to a novel axion insulator state with quantized topological magnetoelectric effect (TME), as shown in Figs. 4c-d. Experimentally, such sandwich heterostructure is grown by MBE where the V and Cr dopants are confined at the top and bottom 3 QL respectively, separated by an un-doped $(Bi_xSb_{1-x})_2Te_3$ spacer layer with thickness from 4 QL to 6 QL [42]. Consequently, each surface contributes a half-integer quantization, and when their magnetization alignment is anti-parallel, the chiral edge states are cancelled; instead, the axion insulator state with zero-Hall plateau is observed, as revealed in Fig. 4e.

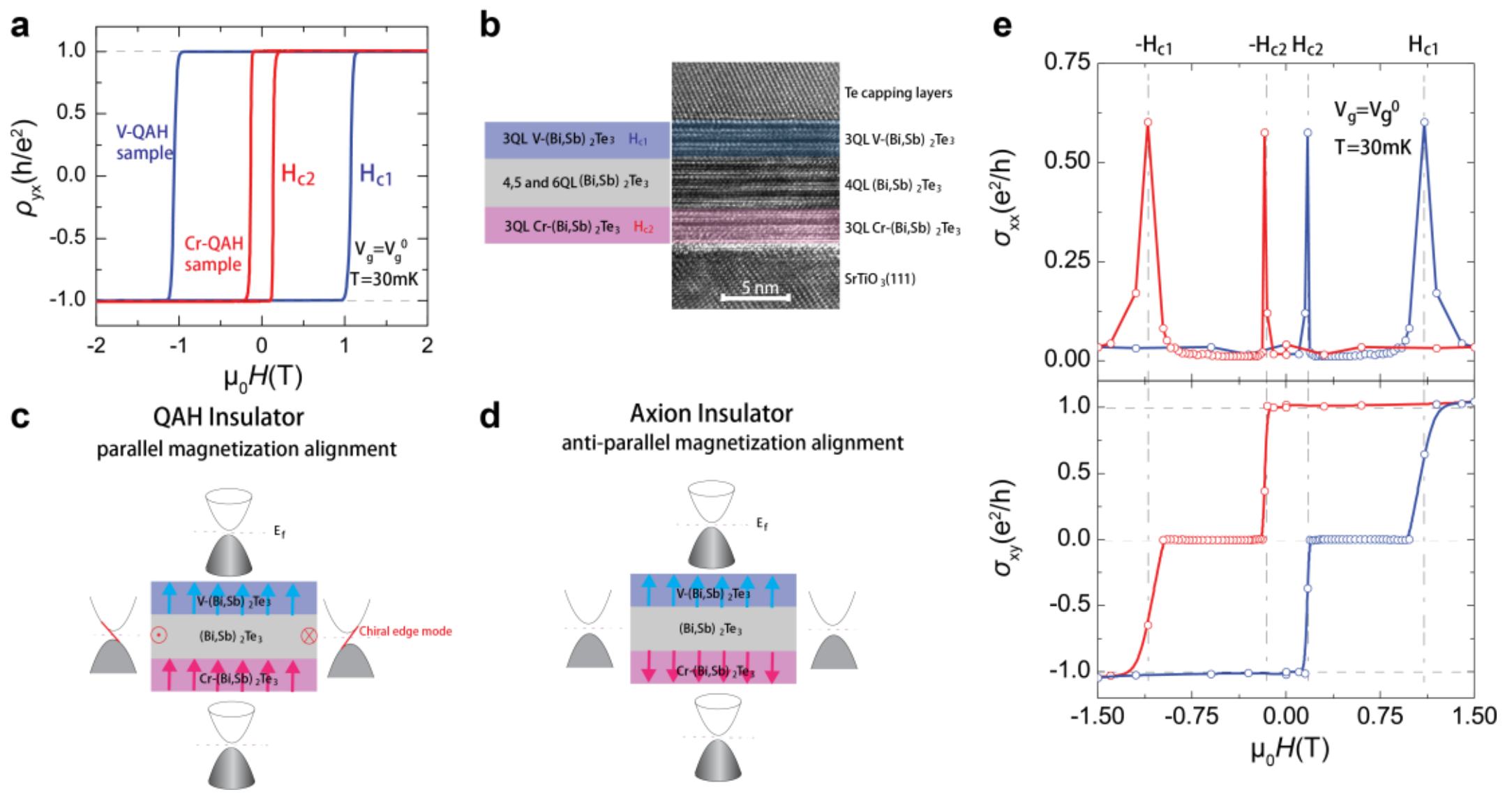

**Figure 4 Realization of the axion insulator state in a 3D V-doped $(Bi,Sb)_2Te_3$/TI/Cr-doped $(Bi,Sb)_2Te_3$ sandwich heterostructure. a**. Signatures of the QAHE in individual Cr- and V-doped $(Bi,Sb)_2Te_3$ films. **b**. Schematic drawing and HRSTEM of the 3 QL V-doped $(Bi,Sb)_2Te_3$/4 QL TI/3QL Cr-doped $(Bi,Sb)_2Te_3$ sandwich heterostructure. **c**. Schematics of a sandwich heterostructure with parallel magnetization alignment surfaces, supporting a 1D chiral edge state. **d**. Schematics of a sandwich heterostructure with antiparallel magnetization alignment surfaces, supporting an axion insulator state. **e**. Experimental observation of the zero-Hall plateau as the consequence of antiparallel magnetization alignment. Reproduced from [42].

In addition to higher-temperature QAH effect and axion insulator state, special design of the magnetic doping profile by modulation-doping method has also enabled us to explore massive Dirac fermion-associated phenomena. Firstly, it is confirmed in the semi-magnetic $(Bi_zSb_{1-z})_2Te_3/Cr_x(Bi_ySb_{1-y})_2Te_3$ bilayer structure, the topological surface states would align the magnetic moments in a distinctive manner owing to their linear energy band dispersion relationship, and the surface-related interaction strength can be either amplified or suppressed, depending on the spatial separation of doped magnetic ions from the top surface [39]. In the meantime, while the QAH edge state on the magnetically-doped surface always contributes to half-integer quantized conductance ($e^2/2h$), the other non-magnetic surface can be driven into the quantum Hall states with discrete Landau levels under high magnetic fields. Based on this scenario, an extraordinary QAH/QHE-hybrid quantum transport behaviors (with filling factor transition from $\nu$=0 to +1 by gate tuning) are observed [43]. More importantly, the concurrent possession of three critical ingredients, namely large SOC, perpendicular magnetic anisotropy, and spatial inversion symmetry-breaking, makes the TI/MTI heterostructures a promising platform for spin-orbit torque-driven magnetization switching, as we will elaborated in Section 3.

Additionally, we should also point out that the precise thickness control by MBE will act as another important tuning parameter in MTIs. In particular, when the MTI film thickness is reduced into the 2D region, the hybridization between the top and bottom surfaces induces a topologically trivial gap that may alter the Chern number, and such parity change plays a critical role for both the realization of metal-to-insulator QAH phase switching [43] and the generations of chiral Majorana fermions [44]. In summary, the functionality of MTIs can be multiplied through the interplay among the band topology, the magnetic orders, and structural engineering via doping profile design and quantum confinement.

## 2. Towards High-$T_C$ Topological Insulators-Based Magnetic Heterostructures

As addressed in Section 1, magnetic doping is an effective way to introduce magnetism, yet all magnetically-doped TI systems inherit low Curie temperatures due to finite *p*-*d* exchange interaction [11-14], and the elevation of Curie temperature with higher doping level (i.e., the highest $T_C \sim 190$ K is obtained in the heavily-doped $Cr_{0.59}Sb_{1.41}Te_3$ [45]) is at the price of degraded sample quality and deteriorated magneto-electric properties. In the meanwhile, the competition between SOC and magnetic coupling strength in a single MTI system might also limit the realization of MTI-featured phenomena towards room temperature. On the contrary, new breakthroughs may emerge by proximity coupling of a high-quality TI layer to a high-$T_C$ ferromagnet. In these MTI heterostructures, the separation of topology and magnetism in different layers not only enables us to optimize each contribution independently, but also enriches the choice of materials and structures in which different physical orders can be better manipulated using all-electrical means.

To implement the proposed high-$T_C$ MTI heterostructures, early attempts have been tried on the growth of TI thin film on ferromagnetic insulator (FMI) substrates/layers (e.g., $Y_3Fe_5O_{12}$ (YIG) and EuS [46-51]), and the interfacial proximity effect has been probed through various techniques. For instance, by comparing with the pure EuS control sample, the Superconducting Quantum Interference Devices (SQUID) data of the $Bi_2Se_3$/EuS heterostructures exhibit enhanced saturation magnetic moment whose origin is likely to be the induced spin-polarization at the interface [52]. From the subsequent polarized neutron reflectometry (PNR) measurements, it is suggested that the induced interficial ferromagnetism could extend ~2nm into the $Bi_2Se_3$ bulk and might persist up to room temperature [49]. Similarly, the magneto-optic Kerr effect (MOKE) results confirm the presence of the FM order up to 130 K in the $Bi_2Se_3$/YIG bilayer system, and the proximity-induced butterfly-shape magnetoresistance (MR) loops are also observed by magneto-transport measurements [48]. However, given that both EuS and YIG are of in-plane magnetic anisotropy, the interfacial $Bi_2Se_3$ surface state can only be gapped by the out-of-plane magnetization at the domain wall regions of FMIs [48]. Accordingly, the resulting weak anomalous Hall effect (AHE) signals without obvious hysteresis loops may obstruct practical spintronics applications.

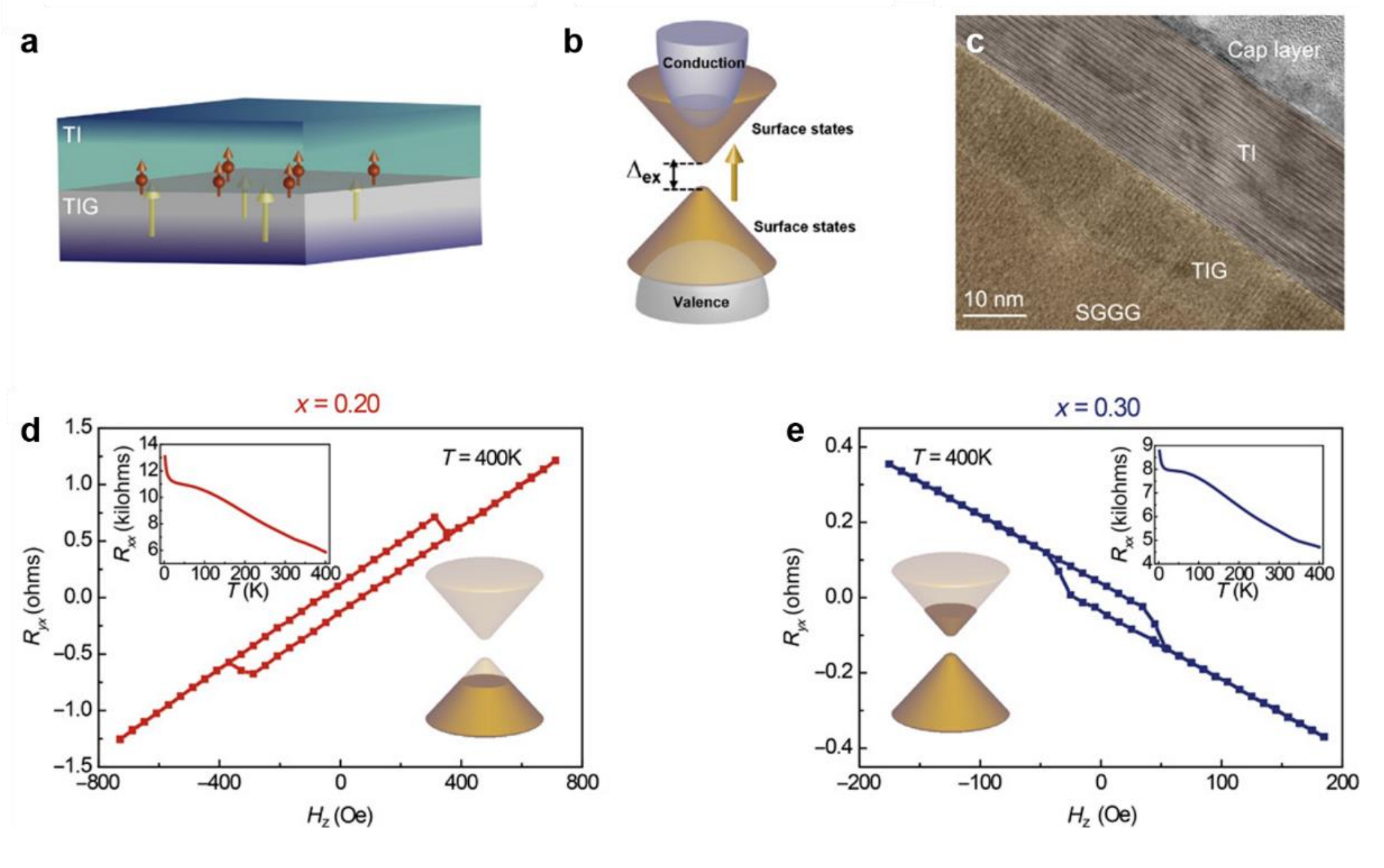

**Figure 5 Proximity-induced ferromagnetism and pronounced AHE at 400 K in the $(Bi_xSb_{1-x})_2Te_3$/TIG/SGGG heterostructure**. **a**. Schematic drawing of proximity coupling between 5QL $(Bi_xSb_{1-x})_2Te_3$ and TIG. **b**. Exchange gap at the charge neutral point of TI surface states induced by broken TRS. **c**. HRTEM image of a TI/TIG (20 QL) bilayer structure. **d-e**. Anomalous Hall traces of the $(Bi_xSb_{1-x})_2Te_3$/TIG (5 QL)

systems for $x$ =0.20 and 0.30, respectively. The upper insets show the corresponding temperature dependence of $R_{xx}$. The lower insets show schematic drawings of the corresponding chemical potential position. Reproduced from [53].

Instead, by achieving the desired perpendicular magnetic anisotropy in the high-quality $Tm_3Fe_5O_{12}$ (TIG) buffer layer grown by pulsed laser deposition (PLD), a robust ferromagnetic phase with record-high $T_C$ > 400 K is been introduced in the adjacent $(Bi_xSb_{1-x})_2Te_3$ thin film [53]. TIG is a rare-earth garnet with ferrimagnetism originating from the antiferromagnetically coupled iron magnetic moments via superexchange interaction. The (111)-oriented substituted gadolinium gallium garnet (SGGG) substrates is deliberately utilized for the epitaxial growth of the TIG layer so that the interface tensile strain exerted by SGGG produces PMA due to the negative magnetostriction constant of TIG (Figs. 5a-c). Consequencely, the TIG layer would drive the interfaical spins towards the direction normal to the TI film plane without any external magnetic field. Accordingly, Figs. 5d-e summarize the marked anomalous Hall effect with well-defined square $R_{xy} - H$ hysteresis loop at 400 K in both $n$-type and $p$-type regimes of this hybrid MTI system.

Interestingly, during the exploration of high-$T_C$ MTI heterostructures, it is noticed that the observed magnetic behaviors may be quite different even for the same TI thin film grown on the same FMI substrates. For example, the reported $T_C$ displays a large disparity range from 20 to 180 K in the $Bi_2Se_3$/YIG heterostructrues [46-48]. It is later unveiled that such sample-to-sample variation mainly stems from the interface quality: the lattice mismatch between TI and FM substrate would inevitably lead to a chalcogenide-rich dead layer during the initial MBE growth stage, which not only jeopardizes the quality of the as-grown film, but also hinders the extension of the magnetic order into the TI layer [48]. Moreover, since the topological feature of the TI surface is vulnerable to the distortion of crystal structure, the existence of quasi-ordered arrays of heavy atoms in some interfacial regions, as well as rotations and tilting between adjacent grains and basal twinning, may also impede the short-range magnetic coupling strength with reduced anomalous Hall resistance (i.e., $< 1\ \Omega$) [54]. Even though a dedicate wet transfer technique is developed to integrate the wafer-scale TI film onto the FM substrate to improve interface quality, yet it can only be applied to limited TI materials (e.g., the KOH used in this recipe can etch other Te-based TIs) and the TI surface would easily get oxidized during the transfer process [55].

Alternatively, the *in situ* hetero-epitaxial growth of the TI/FM heterostructures with similar crystalline configuration may guarantee the formation of the atomically-sharp interface in an ultra-high vacuum environment. The improvement of interface quality in turns gives rise to a more pronounced AHE where large $R_{xy} \sim 120\ \Omega$ is achieved in the MBE-grown $(Bi,Sb)_2(Te,Se)_3/(Ga,Mn)As$ heterostructures [56], and such value is further enlarged to $k\Omega$ in the $Cr_2Ge_2Te_6/TI/Cr_2Ge_2T_6$ sandwich structure at low temperatures, as highlighted in Figs. 6a-b [57]. Furthermore, owning to the giant spin susceptibility of the band-inverted surface states at the well-defined $(Bi_xSb_{1-x})_2Te_3/MnTe$ interface, the uncompensated for $Mn^{3+}$ *d*-orbital electrons can effectively couple with the local TI band electrons to form a new magnetic order whose strength depend on the specific band structure (i.e., Berry curvature) at the Fermi level [54].

Under such condition, by adopting the counter-doping method, the Fermi level position is tuned across the Dirac point with varied Bi-to-Sb ratio, and the polarity of the induced anomalous Hall hysteresis loop is switched, as manifested in Figs. 6d-e. The advantage of tailoring both the amplitude and sign of the AHE response in the above MTI heterostructures may facilitate the SOT-related all-electric device applications.

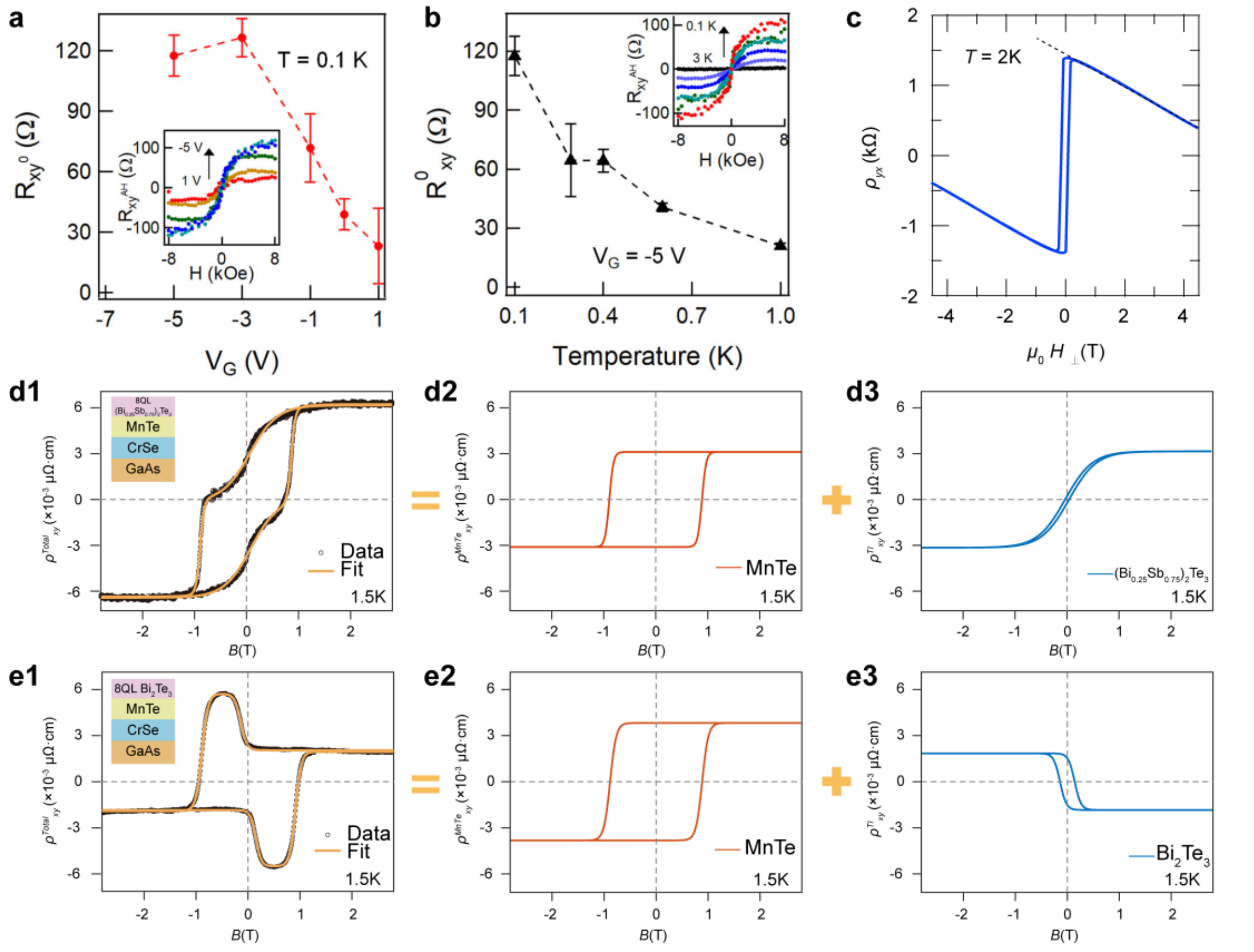

**Figure 6 The manipulation of the AHE response in MTI heterostructures. a-b**. Gate-voltage and temperature dependence of the anomalous Hall resistance $R_{xy}$ at $T$ = 0.1 K. Reproduced from [56]. **c**. Large anomalous Hall signal observed in the CrGeTe/TI/CrGeTe system. Reproduced from [57]. **d-e**. Hybrid AHE results with opposite polarities of the *p*-type 16 QL $(Bi_{0.25}Sb_{0.75})_2Te_3$/MnTe and *n*-type 16 QL $Bi_2Te_3$/MnTe samples, respectively. Reproduced from [54].

In addition to TI/FM combination, similar proximity-induced surface state magnetization can also occur at the interface between TIs and antiferromagnets. Although antiferromagnetic materials do not exhibit macroscopic magnetization, in principle, short-range interfacial exchange coupling to an uncompensated antiferromagnetic plane can locally magnetize the topological surface states if an atomically sharp interface is achieved (Fig. 7b). In this regard, one antiferromagnetic material CrSb, with the Néel temperature ($T_N$) up to ~700 K, is chosen to integrate with $(Bi_xSb_{1-x})_2Te_3$ [58, 59]. Given similar lattice constants between the two constituents, one can use them as the building block to construct various types of Cr-doped $(Bi_xSb_{1-x})_2Te_3$/CrSb heterostructures and super-lattices by MBE (Fig. 7c). Strikingly, the CrSb layer is shown to be an efficient interfacial- and interlayer-exchange coupling mediator between the spins inside different Cr-doped $(Bi_xSb_{1-x})_2Te_3$ layers, which subsequently allows a giant enhancement in magnetic ordering (Fig. 7d) and a modification of the composite magnetic order. Besides, the hysteresis AHE signal and magnetic order can survive up to 90 K (i.e., above the liquid nitrogen temperature), hence uveiling new opportunities for high-temperature topological antiferromagnetic spintronics.

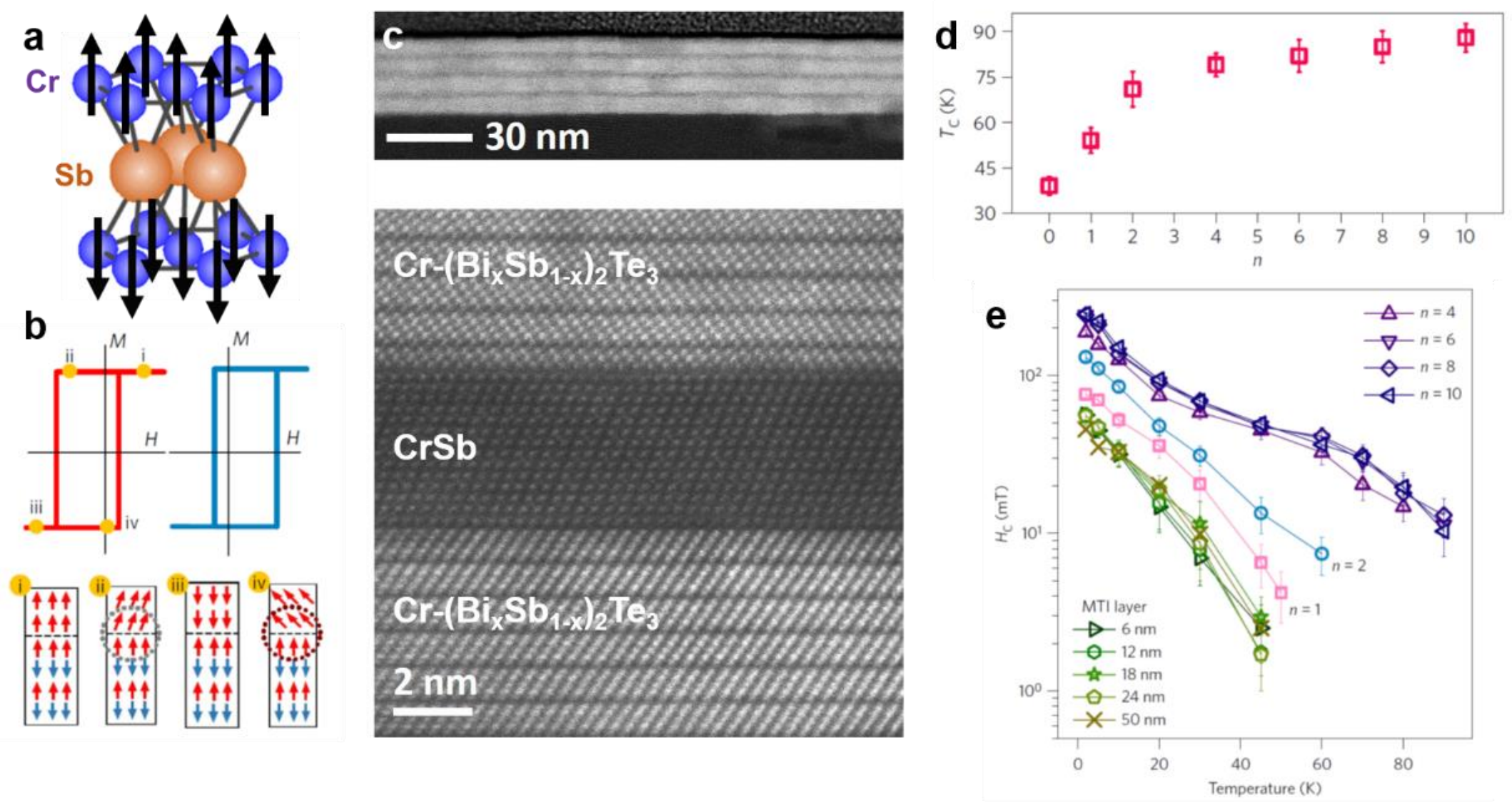

**Figure 7 Tailoring magnetic exchange couplings in magnetic topological insulator/antiferromagnet heterostructures. a**. The spin structures of the AFM-phase CrSb. **b**. Schematic presentation of the atomic moments in the AFM/MTI bilayer structure. **c**. HRSTEM of the CrSb/Cr-doped $(Bi_xSb_{1-x})_2Te_3$ heterostructure. **d**. The enhancement of Curie temperature by structural engineering. **e**. Temperature-dependent coercivity field $H_C$ in different structures. Reproduced from [58].

Recently, by further optimizing the interfacial magnetic proximity effect, the quantum anomalous Hall effect with tunable Chern numbers has also been realized in the emergent MTI heterostructures [60-62]]. Following the concept of magnetic insulator with non-zero Chern number elaborated in Section 1, a novel FMI/TI/FMI sandwich heterostructure of (Zn,Cr)Te/$(Bi,Sb)_2Te_3$/(Zn,Cr)Te is grown by MBE, as illustrated in Fig. 8a [61]. In this hybrid system, it is found that the common Te atoms incorporated in both the TI and FMI layers would permit the topological surface states to deeply extend into the FMI. The resulting strong magnetic exchange interaction at both TI/FMI interface thereafter opens the exchange gap of the TI surface states, and gives rise to the dissipationless chiral edge conduction when all dissipative channels in the (Zn,Cr)Te and $(Bi,Sb)_2Te_3$ bulk layers are suppressed at 30 mK (Fig. 8b). Additionally, the preparation of single-crystalline $Cr_2O_3$ thin film with perpendicular spin-polarized surface by MBE can also introduce highly-efficient interfacial exchange coupling in the Cr-doped $(Bi_xSb_{1-x})_2Te_3/Cr_2O_3$ bilayer sample (Fig. 8c) [60]. Consequently, the induced exchange bias effect provides another degree of freedom to manipulate the QAH states in the TI-based magnetic heterostructures.

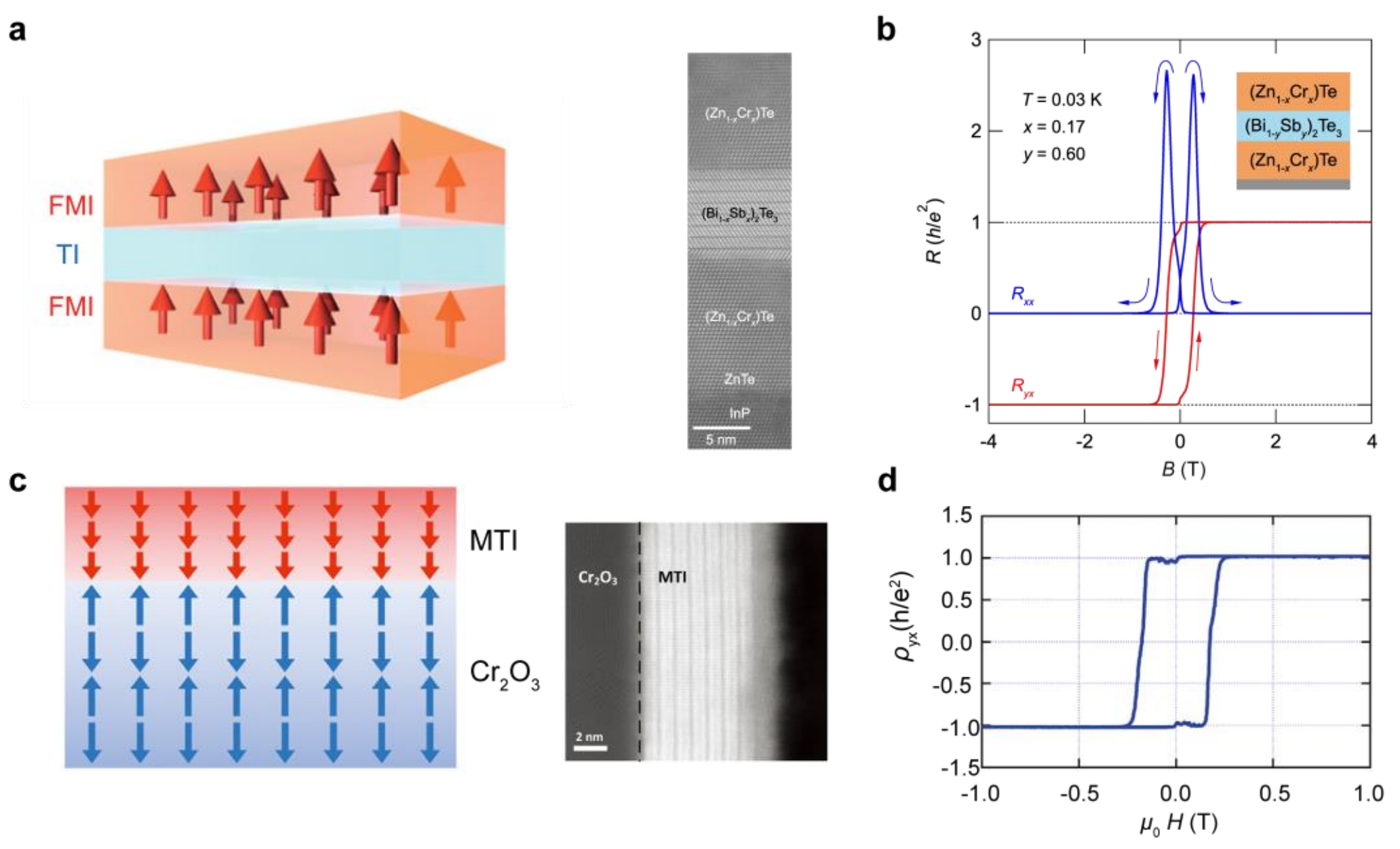

**Figure 8 Observation of QAH effect in the TI-based magnetic heterostructures.** **a**. Schematic drawing of an FMI/TI/FMI sandwich heterostructure and the relevant cross-sectional HAADF-STEM image of the MBE-grown $(Zn,Cr)Te/(Bi,Sb)_2Te_3$ /(Zn,Cr)Te sandwich heterostructure. **b**. Magnetic field dependence of $R_{xx}$ (blue) and $R_{xy}$ (red) manifest the realization of the QAH states at $T = 0.03$ K. Reproduced from [61]. **c**. Spin alignment near the MTI/$Cr_2O_3$ interface and corresponding HAADF-STEM of the MTI/$Cr_2O_3$ heterostructure. **d**. Quantized Hall resistance observed in the MTI/$Cr_2O_3$ sample at 20 mK. Reproduced from [60].

## 3. TI-Based Magnetic Heterostructures for Spintronics Applications

The essence to realize low-power spintronics devices lies in the manipulation of the spin states through electric fields [63]. Accordingly, the TI-based magnetic heterostructures take full advantages of large SOC and tunable magnetic coupling within one system, hence enabling the spin accumulation and precession in an electrically controllable manner. Particularly, the non-trivial band inversion caused by the giant SOC in TIs endorses the spin-momentum locking feature, that is the spin polarization of the surface/edge state carrier is tightly perpendicular to the charge current direction [1, 16]. In the meantime, unlike heavy-metal counterparts, the semi-insulating nature of the bulk TI band structure warrants effective electric-field control of the SOC-associated magnetic/spin properties of the system. In addition, owing to hexagonal warping of the Dirac cone, the surface spin current would carry both in-plane and out-of-plane components, which in turns may generate unique magneto-electric effects and exotic spin textures in the adjacent magnetic layer [64-67]. Finally, the versatile band engineering and interfacial effects in the synthesized heterostructures can further propel the development of MTIs for non-volatile spintronics applications.

In terms of spin-related logic computing, the TI surface states can be used as the spin-signal conduction channel and relevant spin-injection measurements have been conducted in the TI/FM heterostructures [68-70]. By rotating the magnetization of the FM contacts with applied in-plane magnetic field, the channel resistance is switched and the polarity of the resistance hysteresis loop is determined by the electric current direction. Besides, the amplitude of the two-resistance state can be tuned by the applied gate voltage, affirming the operational principle of spin field-effect transistor (FET) prototype. However, the calculated effective spin polarization is much lower than the theoretical expectation, mainly owning to the spin-impedance mismatch and scattering at the TI/FM interface. Moreover, since such TI-based spin-polarized FET device modulate spin of the itinerant electron, the increased bulk conduction ratio at higher temperatures reduces the spin efficiency, therefore disabling the device function above 10 K.

On the other hand, the design of non-volatile magnetic memory devices based on collective spins and correlated magnetic behaviors may be more suitable for room-temperature operation. In fact, it has been observed in the heavy metals (e.g., Pt, Ta, W) with strong SOC, the applied in-plane charge current can generate pure spin current along the transverse direction which arises from either the spin Hall effect and/or interfacial Rashba-Edeistein effect [71-73]. Such spin current, once injected into the adjacent magnetic layer (e.g., Co, CoFeB), can apply strong spin-orbit torque to control the magnetization dynamics of the hybrid system [71]. Compared with the conventional HM/FM films, the charge-to-spin conversion as well as SOT strength is expected to be boosted in the TI-based magnetic heterostructures because the giant SOC and spin-momentum locking of the TI surface states can help to enhance the interfacial spin accumulation.

Motivated by the mechanism discussed above, a spin torque ferromagnetic resonance

(ST-FMR) technique is used to determine the SOT magnitude of TI in the $Bi_2Se_3$/Py bilayer sample (Fig. 9a) [74]. Under the ferromagnetic resonance condition with matched microwave frequency and in-plane magnetic field, the oscillating current-induced SOT in the $Bi_2Se_3$ layer causes the permalloy magnetization to precess, and the resulting ST-FMR spectrum reveals a large in-plane damping-like torque at room temperature (i.e., the symmetric component of the resonance line shape displayed in Figs. 9b-c). Remarkably, although the spin conductivity of the TI/FM sample is comparable to HM/FM systems, the extracted spin torque ratio of $Bi_2Se_3$ (2.0 – 3.5) is up to 40 times larger than that of Pt/W/Ta at 300 K [71, 75, 76]. In the follow-up investigations, different techniques have been performed to elucidate the giant SOT origin and spin-charge conversion mechanism in assorted TI/FM heterostructures including spin pumping, unidirectional magnetoresistance, second harmonic magnetometry, spin Seebeck effect, and non-local spin valve or tunnel junction measurements [65, 68, 69, 77-84].

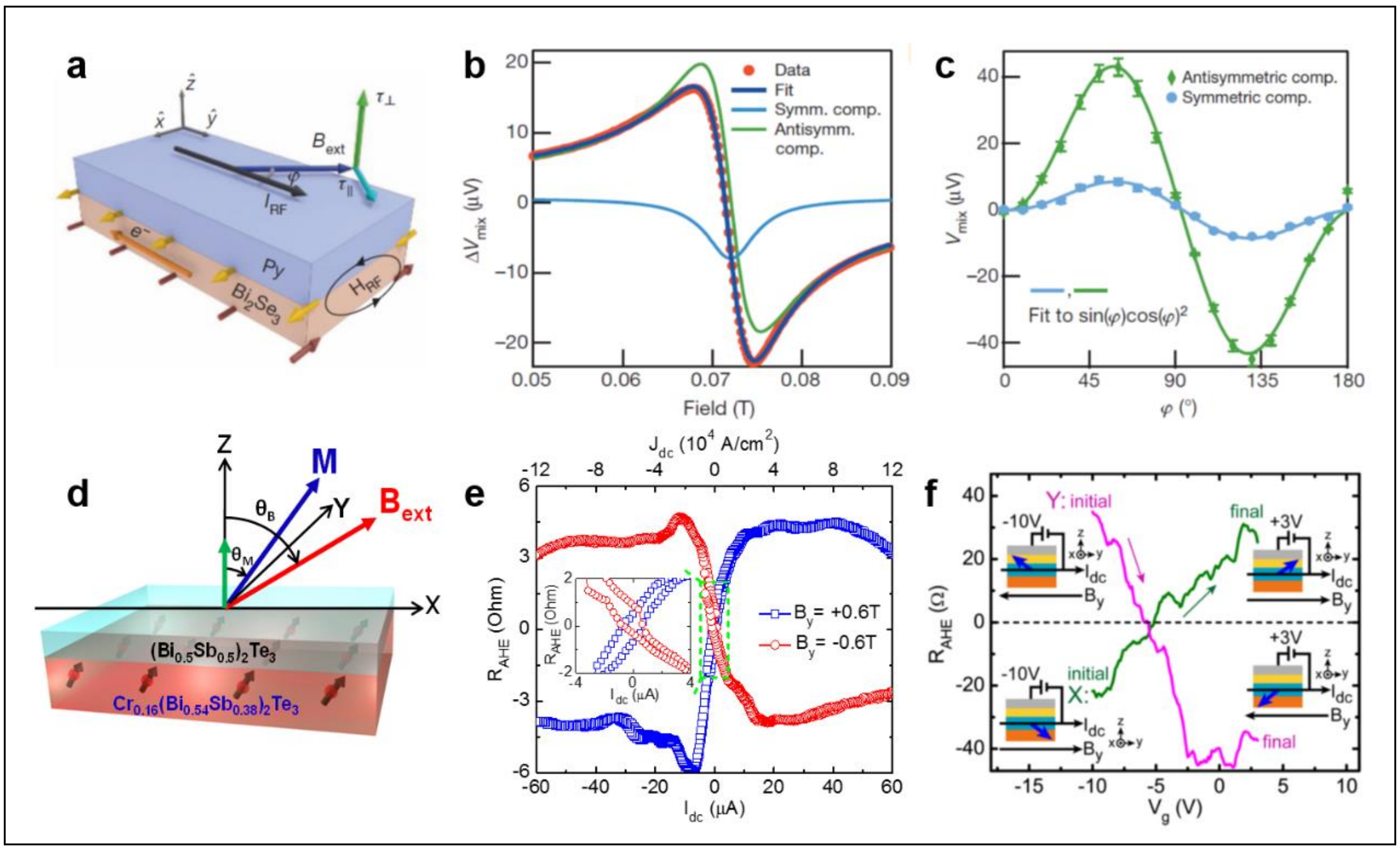

**Figure 9 Giant spin orbit torque and magnetization switching in TI-based magnetic heterostructures. a**. Schematic diagram of the $Bi_2Se_3$/Py bilayer structure used in the ST-FMR measurement. The yellow and red arrows denote spin moment directions. **b**. Measured ST-FMR resonance at room temperature with microwave frequency of 8 GHz for an 8 nm $Bi_2Se_3$/16 nm permalloty sample. **c**. Measured dependence on the magnetic field angle $\varphi$ for the symmetric and antisymmetric resonance components for the TI/FM thin film. Reproduced from [74]. **d**. Illustration of the bilayer TI/MTI heterostructures with applied magnetic field. **e**. Current-induced magnetization switching in the TI/MTI Hall bar device at 1.9 K in the presence of a constant in-plane magnetic field. Inset: zoom-in scale to show the hysteresis windows. Reproduced from [10]. **f**. Electric-field control of the SOT-driven magnetization switching in a top-gated $Cr_{0.16}(Bi_{0.5}Sb_{0.42})_2Te_3$ thin film with inversion symmetry breaking. Reproduced from [81].

Most intriguingly, the magnetization switching due to the current-induced SOT is demonstrated in the modulation-doped $(Bi_{0.5}Sb_{0.5})_2Te_3/(Cr_{0.08}Bi_{0.54}Sb_{0.38})_2Te_3$ heterostructures [80]. In this TI/MTI bilayer structure, the TI layer provides the pronounced spin-polarized current from the topologically nontrivial surface states, while the MTI layer acts as a robust ferromagnet with perpendicular anisotropy. As illustrated in Fig. 9d, when the DC current ($I_{dc}$) and in-plane magnetic field ($B_x$) are applied along the +$x$-axis, the effective spin-orbit field ($B_{SO}$), which points along the tangential -$\theta$ direction, can drive the switch of the MTI magnetization (Fig. 9e). Strikingly, it is seen that the critical current density (i.e., below $10^5$ A/cm$^2$) is about three orders of magnitude smaller than that required for heavy metallic systems, confirming that the current-induced SOT in the TI/MTI heterostructures is extremely efficient. With further fabricating the MTI-based top-gate FET device, the electric field manages to modulate the SOT strength (i.e., which is closely correlated with the spin-polarized surface current) by a factor of four [10]. Based on the SOT phase diagram, the magnetization switching by scanning gate voltage with constant current and in-plane magnetic field is also realized, as shown in Fig. 9f.

In pursuit of the distinctive SOT-related effects for low-power spintronics applications towards room temperature, replacing the magnetically-doped MTI in the above example by a high-$T_C$ magnetic layer is proved to be a reliable route. Generally, to validate the SOT-driven magnetization switching via magneto-transport measurement, it is required that the neighboring FM layer should have perpendicular magnetic anisotropy. However, the magnetic anisotropy of the commonly used high-$T_C$ materials (e.g., CoFeB and Py) strongly relies on the interfacial conditions, and it usually favors an in-plane orientation when the FM thin film is directly grown on the TI layer [85-88]. To address the challenge, one can insert a light metal spacer (i.e., ~ 2nm Mo or Ti) to form the TI/Ti(Mo)/CoFeB/MgO multilayer structures so that interfacial PMA can be established while keeping the spin current in the TI layer (Fig. 10a) [86, 87]. Alternatively, by depositing bulk PMA material (e.g., ferrimagnetic alloy $Co_xTb_{1-x}$ with optimized Co-to-Tb ratio), room-temperature SOT switching has also been observed in the $Bi_2Se_3$/CoTb/$SiN_x$ stacks [88]. More recently, by utilizing the MOKE microscopy, the SOT-induced magnetization switching dynamics can also be visualized in the $Bi_2Se_3$/Py stacks with in-plane anisotropy at room temperature, as shown in Fig. 10c [85].

To further improve the SOT efficiency in the TI/FM heterostructures, several structural engineering strategies have been adopted [85-87, 89]. Given the spin-momentum locking nature, the topological surface states are expected to generate more pronounced spin current than the TI bulk states. Accordingly, we can either tune the Fermi level towards the Dirac point by adjusting the Bi-to-Sb ratio in the $(Bi_{1-x}Sb_x)_2Te_3$ layer or reduce the TI film thickness so that bulk dissipative channels are suppressed [90, 91]. As confirmed by Figs. 10b and 10d, both methods can significantly enhance the SOT efficiency which corresponds to the surface states-dominated conduction scenario. In addition, it is also justified that the *in-situ* preparation of both the conductive TI and magnetic layers in one growth system (e.g., BiSb/MnGa by MBE [92] and $Bi_xSe_{1-x}$/CoFeB by magnetron sputtering [89]) can yield better interface quality and high SOT strength at room temperature.

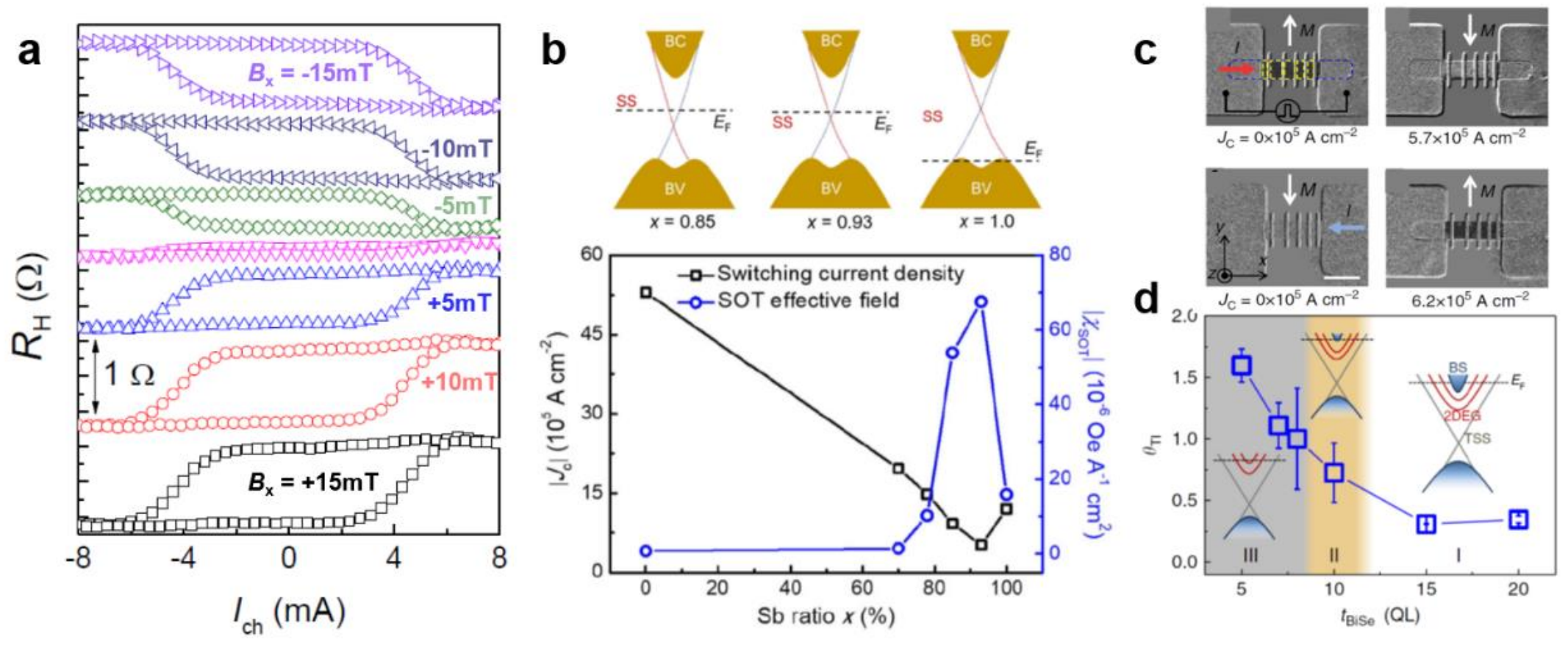

**Figure 10 SOT-driven magnetization switching at room temperature and the SOT efficiency manipulation in TI/FM heterostructures. a**. Room-temperature

SOT-driven magnetization switching in the $(BiSb)_2Te_3$(6 nm)/Mo(2 nm)/CoFeB(1.02 nm) sample. Reproduced from [87]. **b**. Switching current density and SOT-induced effective field as a function of Sb ratio in the $(Bi_{1-x}Sb_x)_2Te_3$ material. Reproduced from [86]. **c**. Visualization of the SOT switching dynamics in the $Bi_2Se_3$/Py at zero magnetic field and room temperature by MOKE. **d**. The SOT efficiency as a function of $Bi_2Se_3$ thickness at room temperature. Reproduced from [85].

Equivalently, the strong SOC combined with inversion symmetry breaking in TI-based magnetic heterostructures can also provide the non-collinear Dzyaloshinskii–Moriya interaction (DMI) at the interface, hence facilitating the formation of non-trivial topological spin textures (e.g., magnetic skyrmions) and exotic topological/geometric Hall effects [93, 94]. Experimentally, the trace of magnetic skyrmions can be found among modulation-doped MTI, MTI/FM and MTI/AFM heterostructures [95-97], and their unique properties has been characterized by magneto-transport, X-ray photoemission electron microscopy (XPEEM) and Scanning Transmission X-ray microscopy (STXM), as summarized in Fig. 11. Preliminary data have also implied that the mediation of DMI and PMA via material composition helps to stabilize the Néel-type skyrmions, and the real-space Berry curvature from their sophisticated spin configuration might be responsible for the appearance of topological Hall-like line-shape from the measured anomalous Hall results [95].

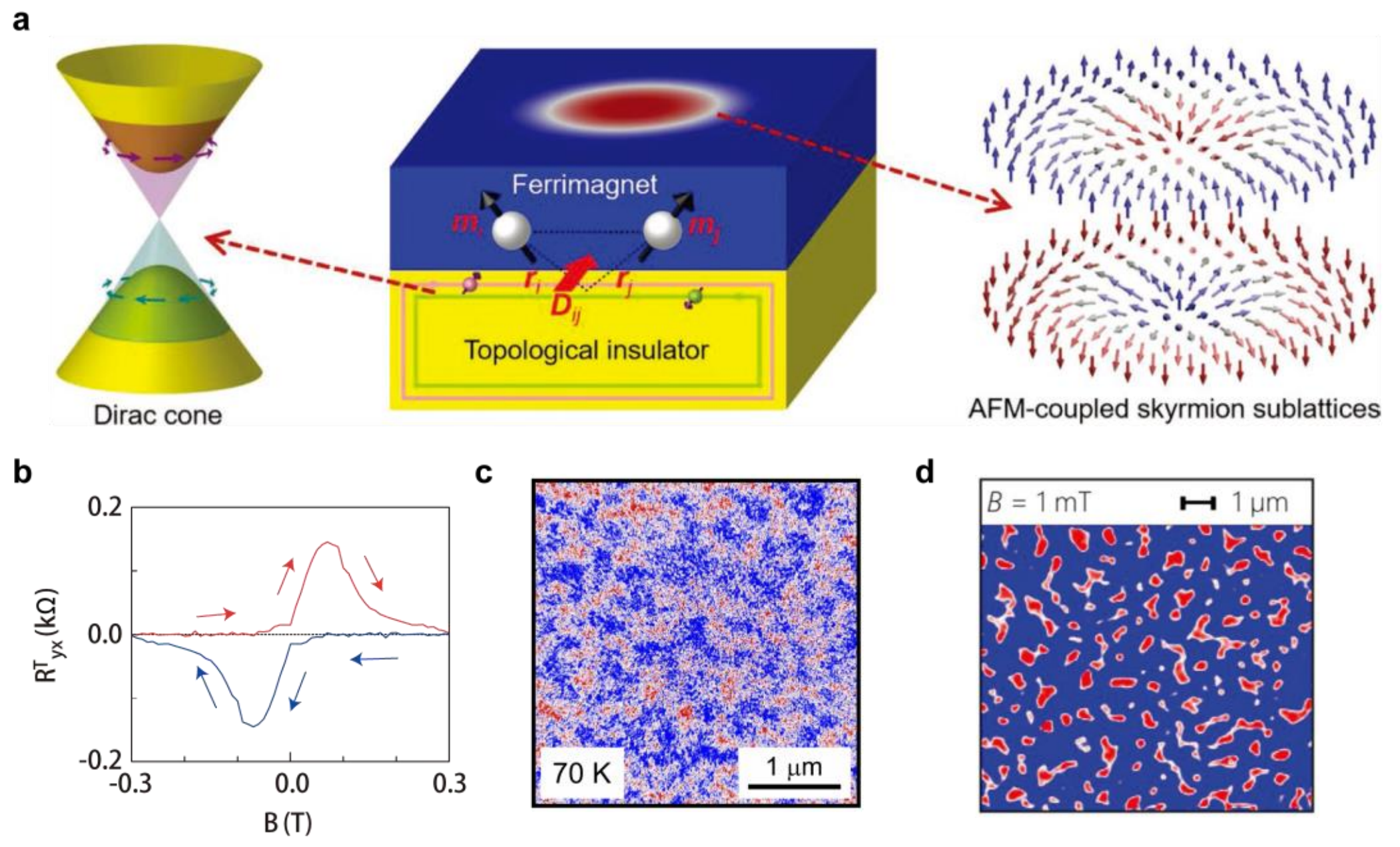

**Figure 11 Exotic spin textures and topological Hall effects in TI-based magnetic heterostructures.** **a**. Schematic of the DMI interaction and skyrmions in TI/ferrimagnet heterostructures. Reproduced from [94]. **b**. Topological Hall effect observed in the $Cr_x(Bi_{1-y}Sb_y)_{2-x}Te_3/(Bi_{1-y}Sb_y)_2Te_3$ heterostructures. Reproduced from [95]. **c**. XPEEM image of the magnetic domain structure in the NiFe/Cr-$Sb_2Te_3$ sample. Reproduced from [96]. **d**. STXM pattern of the skyrmions in the $(Bi,Sb)_2Te_3$/GaFeCo heterostructure. Reproduced from [94].

## 3. Challenges and Opportunities

From the above discussion, it is obvious that TI-based magnetic heterostructures have distinguished themselves from other candidates because of their salient capabilities to incorporate high-$T_C$ magnetism and to tailor different physical orders through effective band engineering and interfacial/low-dimensional effects. To further explore TRS-breaking physics and low-power non-volatile spintronics applications based on such MTI hybrid framework, new opportunities, which co-exist with potential challenges, may emerge from the following aspects.

First of all, even though the heavily-doped surfaces help to enlarge the overall magnetic exchange gap, the modulation-doped MTI samples still suffer spatial randomness issue from both the magnetic dopants as well as the local Bi-to-Sb variation of the host $(Bi,Sb)_2Te_3$ matrix. Meanwhile, since the wavefunction of the topological surface states is strictly localized at the FMI/TI interface, the induced FM interaction can only open a small surface gap [35]. Therefore, to achieve spontaneous uniform magnetization profile, an intrinsic antiferromagnetic topological insulator $MnBi_2Te_4$ has been proposed and QAH effect has recently been observed in the exfoliated sample at 1.4 K (i.e., other experiments have shown that the quantization of $R_{xy}$ can be obtained in similar systems above 77 K, yet with the presence of large magnetic field) [98]. With further structural optimizations (e.g., inserting non-magnetic TI spacer to reduce the interlayer AFM coupling [99]), zero-field QAH states may occur in the $MnBi_2Te_4$-based heterostructures/superlattices at a higher temperature.

Regarding the high-$T_C$ MTI heterostructures, it is shown that the hetero-interface quality plays an indispensable role in determining the magnetic proximity effect and SOT strength. In this regard, it seems that to prepare the entire TI/FM or TI/AFM structures in the same system (e.g., either in the same growth chamber or different MBE/sputtering/PLD chambers inter-connected via ultra-high vacuum transfer apparatus) may be a better option so that atomically sharp interface with well-ordered atom configuration can be realized. Besides, rapid development in both the topological quantum physics and 2D material research fields have greatly expanded the material categories that can be joined together. For example, the discovery of novel 2D FM materials with layered structure may be epitaxially integrated with TI through the van der Waals growth mode; likewise, a multitude of identified Dirac/Weyl semimetals may bring about exotic electronic structures with non-zero Berry curvature to trigger new spin-related effects and device concepts for MTI heterostructure-based multifunctional applications.

**Acknowledgements**

This work is sponsored by the National Key R&D Program of China under the contract number 2017YFB0405704, National Natural Science Foundation of China (Grant No. 61874172, 11904230, 11874070), the Strategic Priority Research Program of Chinese Academy of Sciences (Grant No. XDA18010000), and the Major Project of Shanghai Municipal Science and Technology (Grant No. 2018SHZDZX02). Q.Y. acknowledges the support from the Shanghai Sailing Program (Grant No. 19YF1433200). Q.L.H. acknowledges the supports from the National Natural Science Foundation of China (Grant No. 11874070) and the Strategic Priority Research Program of Chinese Academy of Sciences (Grant No. XDB28000000).